\documentclass{article}

\usepackage{arxiv}

\usepackage[utf8]{inputenc} 
\usepackage[T1]{fontenc}    
\usepackage{hyperref}       
\usepackage{url}            
\usepackage{booktabs}       
\usepackage{amsfonts}       
\usepackage{nicefrac}       
\usepackage{microtype}      
\usepackage{lipsum}
\usepackage{graphicx}
\usepackage{comment}
\usepackage{xcolor}
\usepackage{makecell}
\usepackage{array}
\usepackage{booktabs,chemformula}
\graphicspath{ {./images/} }

\title{T-Grating on Nano-Cavity Array based Refractive Index Sensor}

\author{
 Yasir Fatha Abed \\
  Department of Electrical and Electronic Engineering\\
  Bangladesh University of Engineering and Technology\\
  Dhaka, Bangladesh \\
   \And
 Md Asif Hossain Bhuiyan \\
  Department of Electrical and Electronic Engineering\\
  Bangladesh University of Engineering and Technology\\
  Dhaka, Bangladesh \\
  \And
 Sajid Muhaimin Choudhury \\
  Department of Electrical and Electronic Engineering\\
  Bangladesh University of Engineering and Technology\\
  Dhaka, Bangladesh \\
  \texttt{Corresponding author: sajid@eee.buet.ac.bd} \\
}

\begin{document}
\maketitle
\begin{abstract}
We report a refractive index sensor comprising of unique T grating on top of periodic nano-cavities. The sensor has two resonant modes sensitive to different regions of the structure with low inter-region interference, hence allows simultaneous detection of two different analytes or more accurate detection of a single analyte. The sensor also provides a self-referencing feature for a broad range of refractive index, from 1.3 to 1.5. Using the FDTD method, the sensitivities of 801.7 nm RIU\textsuperscript{-1} and 1386.8 nm RIU\textsuperscript{-1} have been recorded for the two modes respectively. The versatility of the structure makes the sensor a prominent candidate for biochemical and other sensing applications. 
\end{abstract}


\section{Introduction}
Surface plasmon polaritons (SPPs) involve non-radiative evanescent surface waves excited by the coherent and collective oscillations of the conduction electrons on a metal surface at the metal-dielectric interface \cite{Fano1941}. This oscillation of conduction band electrons generated by the interaction between photons and surface electrons is often referred to as the surface plasmons (SPs). SPs can be localized or propagating in nature \cite{Stewart2008}. Localized surface plasmons (LSPs) are non-propagating plasmonic motions trapped within conductive subwavelength nano-particles (NPs) while propagating surface plasmons (PSPs) travels along the metal-dielectric interface of a nano-structure \cite{Chu2009}. The coupling between photons and surface electrons, known as surface plasmon resonance (SPR), is exceedingly dependent on the dielectric environment surrounding the metal surface \cite{Nazem2020}, size and geometry of the nano-structure \cite{Jiang2018}, and angle of photon incidence \cite{Paliwal2019}.

Alongside SPR, the incorporation of other resonance modes in a hybrid structure has recently developed a great interest in the field of photonics. These structures supporting multi resonance mode like SPR and (i) Surface Enhanced Raman Scattering (SERS) \cite{Lee2018}, (ii) Guided Mode Resonance (GMR) \cite{Fannin2017}, (iii) Fabry-Perot (FP) \cite{Zhu2017} resonances have been investigated previously. Among these resonances, FP-like resonance is very popular due to its narrowband application \cite{Rana2020}. In FP nano-plasmonic cavities, the photon is trapped between two parallel mirrors constituting standing waves and selective wavelengths create resonance depending on the length of the cavity \cite{Kogelnik1966}. As the optical path inside the cavity is dependent on the refractive index of the dielectric material inside \cite{DellaValle2016}, the resonating wavelength is also very sensitive to the refractive index of the dielectric material \cite{Breeden1969}.

Over the past decade, a strong interest in plasmonic structures has been grown as extensive research has been conducted on perfect absorber \cite{ChouChau2019,askari2021near,zhang2021dual}, laser \cite{Oulton2009,jin2020phase,cui2020plasmonic}, label free biosensing \cite{Nath2004,liang2019self,ciminelli2019integrated}, optical filter\cite{Lee2015_filter}, nano-resonator\cite{Yan2017}, short range data transfer in photonic integrated circuits (PICs) \cite{Kim2011}, demultiplexer \cite{Forati2013}, buffer \cite{Yan2017}, polarizer \cite{8125773} and refractive index sensor\cite{Sarker2020,Cen2018,Zhang2018,wang2019theoretical,fannin2017properties,8936420,golfazani2020analytical}. In SPR based refractive index (RI) sensors, phase matching condition for resonance can be satisfied by prism coupling or grating coupling \cite{Maier2007}. In prism and grating coupled configurations, different wavelengths at a particular angle (spectral scheme) or the same wavelength at different angles (angular scheme) can be coupled with resonant modes depending on the refractive indices of the  analyte \cite{Lee2015,sarker2020structurally}. In recent years, many spectral scheme based one dimensional (1D) grating coupled highly sensitive refractive index sensors have been proposed \cite{Afsheen2019,Sakib2018,zhu2020plasmonic,su2021self}. The major advantage of this configuration is the miniaturization of the structure and it has been possible because of the capability of sub-wavelength grating formation by advanced lithography process \cite{koch1995grating}. Real-time monitoring of the variation of concentration of an analyte or even multiple binding events of biomolecules on the grating surface is realizable by grating coupled configuration \cite{Darwish2010,shrivastav2021comprehensive,agrawal2021non,li2021digital}.

Unlike a single analyte based refractive index sensor, the research area of incorporating both SPR and FP-like resonances in a single structure for simultaneous multiple analyte detection has been less explored. Previously, an optical fiber based refractive index sensor was suggested for simultaneous detection of maximum three analytes but with a limited refractive index sensing window of 1.33$\sim$1.35 and a high inter-channel interference which results in a more erroneous detection system \cite{Tabassum2020}. Another reported work proposed a suspended core fiber based multi-analyte sensor with loss spectra interrogation scheme where the difference between refractive indices of the analytes had to be at least 0.03 unit for a better sensing performance which in turn shrunk the RI performance window \cite{Gomez-Cardona2018}. Some previously reported self-referencing RI sensors have the same problem with inter-region interference on the individual performance of single mode deteriorating their self-referencing performance \cite{Sharma2019,Abutoama2015}.

In this paper, we propose a novel plasmonic refractive index sensor with a unique T-shaped grating on top of periodic rectangular cavities. The T-shaped grating and nano-cavities create two active regions enabling simultaneous detection of two different analytes. The grating structure introduces two L-shaped cavities on top of each nano-cavity which entraps incident photon resulting in a better coupling. Two individual resonant conditions couple electromagnetic (EM) field at different regions of the structure at different phase matching conditions allowing the two resonant modes to perform with significantly less inter-region interference than most of the reported multi-analyte and self-referencing bio-sensors.

The numerically calculated sensitivities of SPR and FP-like resonant modes are 801.7 nm RIU\textsuperscript{-1} and 1386.8 nm RIU\textsuperscript{-1} respectively. The inter-region interference adds 5.8 nm RIU\textsuperscript{-1} and 0 nm RIU\textsuperscript{-1} induced sensitivity to the resonant modes respectively between 1.33 to 1.37 RI range which allows performing more accurate multi-analyte detection for relatively broader RI range of interest than the previously reported multi-analyte sensors. For the self-referencing scheme, the sensitivities of the active mode and the reference mode are 800 nm RIU\textsuperscript{-1} and 2.82 nm RIU\textsuperscript{-1} respectively considering silicon dioxide as reference.

Furthermore, the capability and usability of the sensor for simultaneous detection of multi-analyte and self-referencing modes have been briefly analyzed and verified.

\begin{figure*}[!h]
\centering
\includegraphics[clip=true,trim= 0 0 0 0,width=0.95\textwidth]{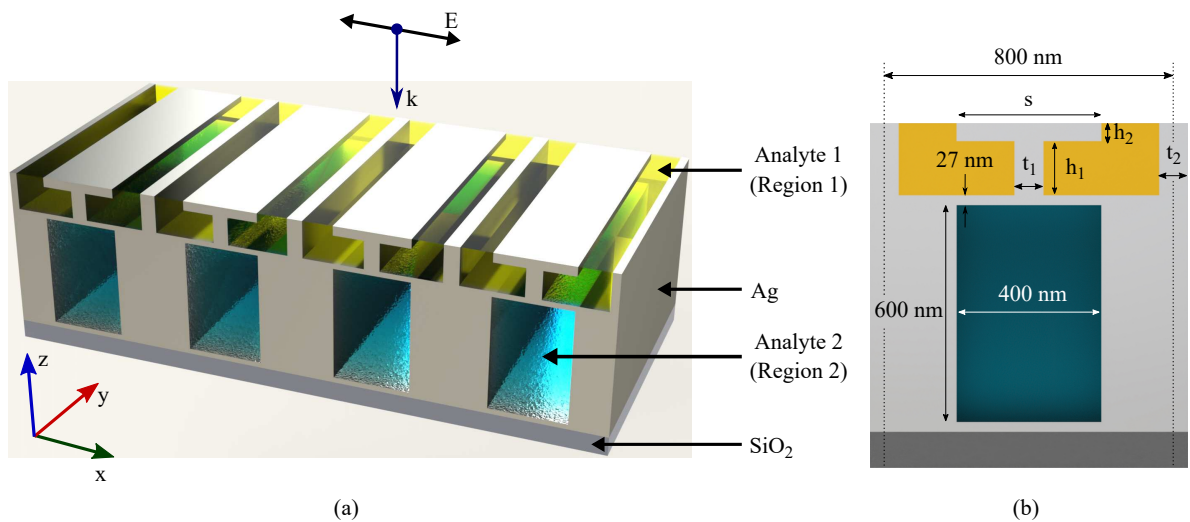}
\caption{\label{fig:sensor}(a) 3D schematic illustration of four unit cells of the proposed plasmonic sensor. (b) 2D front view of one unit cell (depicted in dotted line) of the proposed sensor illustrating different structural parameters.}
\end{figure*}

\section{Sensor Design}

The 3D and 2D schematic view of the proposed sensor is shown in fig. \ref{fig:sensor}(a) and \ref{fig:sensor}(b), respectively. The sensor comprises of T-shaped 1D grating on top of periodic nano-cavities embedded into bulk Ag body on SiO\textsubscript{2} substrate. These nano-cavities of width 400 nm and height 600 nm defining region 2 will house one analyte. On the upper side of the structure, horizontal Ag slab of span s and thickness h\textsubscript{2} is placed upon vertical Ag slab of height h\textsubscript{1} and thickness t\textsubscript{1} to resemble the letter T introducing two L-shaped cavities under the horizontal slab. The reason behind the inclusion of a cavity is that it entraps incident photons under the horizontal silver slab enabling low loss photon-electron interaction and thus resulting in strong SPP coupling in the nano-cavity region \cite{Bisht2019}. The T-shaped structure is then bounded by Ag walls of thickness t\textsubscript{2}. This outer part of the sensor depicted by region 1 can be submerged into another analyte which can be the same or different from the previous one. The period of the structure is set to 800nm. Other structural parameters are set to be variable and the effects of these parameters are to be studied in the later section. A silver layer which thickness is less than the skin depth, separates the two distinctive analyte regions so that the injected electromagnetic wave can penetrate the layer and interact with both of the analyte regions  \cite{Homola2006}.

\begin{figure}[!h]
\centering\includegraphics[clip=true,trim=0 0 0 0, width=0.8\textwidth]{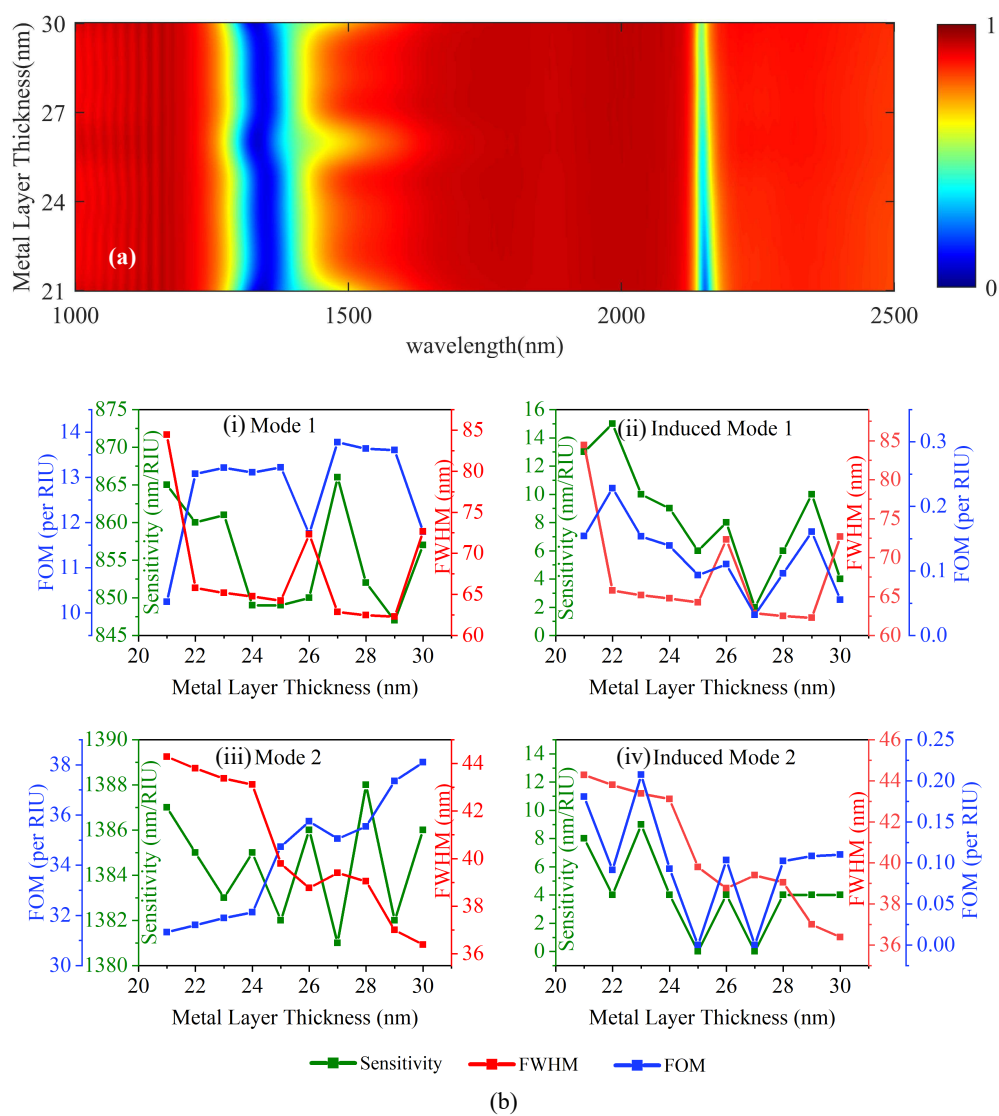}
\caption{\label{fig:MLT} \textcolor{black}{(a) Reflection spectrum for different Ag layer thickness values from 21 nm to 30 nm. Mode 1 and mode 2 resonance do not exhibit any significant spectral shift with the increase of the thickness. (b) (i) Individual and (ii) induced sensing performance of mode 1 and (iii) individual and (iv) induced sensing performance of mode 2 with different metal layer thicknesses. Overall better performance can be achieved for the value of 27nm.}}
\end{figure}
To determine the optimum thickness of the Ag layer in the sensor, the reflection spectrum for various Ag layer thickness is plotted in fig. \ref{fig:MLT}(a). Notably, with the increase in the metal layer thickness, the resonant modes do not show any significant spectral shift. But the mode 2 resonance weakens due to the increase of the Ag layer thickness which is expected because of the attenuation of EM field in Ag \cite{bouhelier2001plasmon}. The individual and induced sensing performance do not show any specific trend for various thickness of Ag layer as depicted in fig. \ref{fig:MLT}(b)(i-iv). However, for 27 nm thickness, overall better performance can be seen. So the optimum thickness of the Ag layer was determined as 27 nm. The structure is illuminated from the above with a transverse magnetic (TM) polarized plane wave at normal incidence \textcolor{black}{along the negative z axis.}  

\textcolor{black}{The proposed structure can be fabricated by employing the standard top-down method of nanofabrication \cite{Yousuf2018,sanchez2021suspended}. First of all, 630 nm of a bulk silver layer can be grown/deposited on a SiO\textsubscript{2} implementing physical vapor deposition technique. Then periodic deep trenches of 600 nm are formed by lithography technique and utilizing reactive ion etching (RIE) of the bulk silver layer and hence, region 2 can be formed. Afterward, the trench region can be filled with a sacrificial photoresist layer and on top of it, a 27 nm thick silver layer can be deposited \cite{nejat2020multi}. Then by employing patterning and deposition techniques available in the standard top-down fabrication process, the aforementioned grating structure can be fabricated. Lastly, the sacrificial photoresist can be removed using acetone and the proposed structure can be obtained.}

\section{Analysis methodology}
For a grating coupled configuration, the phase matching relation given by Eq. (\ref{eq:grating}) characterizes the SPP excitation at the grating-dielectric interface \cite{seo2017grating}.
\begin{equation}\label{eq:grating}
   n_a\sin{\theta_{inc}} + m\frac{\lambda}{\Lambda} = \pm Re(\sqrt{\frac{\epsilon_mn_a^2}{\epsilon_m+n_a^2}})
\end{equation}
Here, $\Lambda$ is the grating period, $\lambda$ is the wavelength of the light, $\theta_{inc}$ is the angle of the incidence and $n_a$ is the refractive index of the dielectric medium. From the phase-matching condition, it is evident that the refractive index of the dielectric has some direct influence on resonant wavelength manipulation. This specific property can be utilized to construct an SPP based refractive index sensor.

For a plasmonic refractive index sensor, two types of interrogation schemes are used: angular interrogation and spectral interrogation \cite{homola1997sensitivity}. In the spectral interrogation scheme for a specific change of refractive index of the surrounding material, there is a certain shift of resonant wavelength in the reflection spectrum. The ratio of the shift of resonance wavelength and the change in RI is denoted by sensitivity \cite{Galopin2010}.
\begin{equation}
  S=\frac{\delta\lambda}{\delta n}  
\end{equation}
The selectivity of the sensor is determined by the ratio of sensitivity to the full-width at half-maximum (FWHM) termed as figure of merit (FOM) \cite{Yousuf2018} which characterizes the overall sensing performance. 
\begin{equation} \label{eq:fom}
  FOM=\frac{S}{FWHM}  
\end{equation}

In this work, the finite difference time domain (FDTD) method is used with proper boundary conditions to model the sensor structure and analyze the electromagnetic response. To ensure the accuracy of the simulation, the mesh size is selected to be as low as 0.2 nm. The Lorentz-Drude model provided by Rakić et al \cite{Rakic1998}  is used to derive the optical parameters of Ag. 

\section{Results and optimization}

\subsection{Reflection spectrum and field profile:}
\begin{figure*}[t]
\centering\includegraphics[clip=true,trim= 0 0 0 0,width=0.85\textwidth]{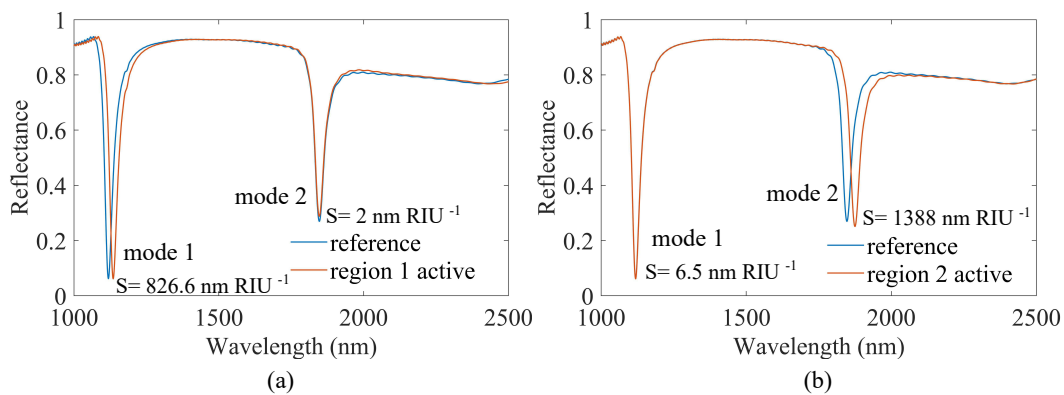}
\caption{\label{fig:spectrum}Reflection spectrum of the sensor while (a) region 1 is active and (b) region 2 is active. In each active region, the corresponding RI undergoes a shift of 0.01 unit.}
\end{figure*}

The reflection spectrum of the sensor plotted in fig. \ref{fig:spectrum} shows two distinctive resonant modes defined as mode 1 and mode 2 \textcolor{black}{where mode 1 corresponds to SPR and can be approximated by using the phase matching condition described by eq. \ref{eq:grating} and mode 2 corresponds to the FP-like resonance which can be estimated by the following resonance condition \cite{li2010long}: }
\textcolor{black}{\begin{equation}
    Re[(\epsilon_m \times k_a - n_a^2)^{1/2}] \times h + \frac{\Delta \phi}{2} = \frac{m\pi}{2}
\end{equation}}
\textcolor{black}{Here k\textsubscript{a} is the wave-vector inside the metal-dielectric-metal (MDM) cavity, $\phi$ is the phase change of the reflected wave and m is an integer denoting the FP mode. Noteworthy, because of the scattering introduced by the grating and propagation loss due to SPP inside the MDM  nano-plasmonic cavity, this resonant condition may not be fully satisfied.} An interesting feature of the proposed structure is that mode 1 and mode 2 are sensitive only to the RI of region 1 and region 2 respectively. It is to be noted that the spectral distance between the modes is high enough to be fairly distinguishable irrespective of the RI of any region. The insignificant spectral shift for one mode due to the RI shift of the neighboring region is defined by the induced sensitivity of that mode. Fig. \ref{fig:spectrum}(a) depicts that for a change in RI of only region 1, mode 1 shows the sensitivity as high as 826.6 nm RIU\textsuperscript{-1} while the induced sensitivity of mode 2 is only 2 nm RIU\textsuperscript{-1}. As plotted in fig. \ref{fig:spectrum}(b), mode 2 exhibits the sensitivity as high as 1388 nm RIU\textsuperscript{-1} while the induced sensitivity of mode 1 is as low as 6.5 nm RIU\textsuperscript{-1} for a change in RI of region 2 alone. Hence a simultaneous spectral interrogation for two different analytes or more reliable sensing of a single analyte or even self-referencing of the structure is possible by employing these two mutually independent resonant modes.  

\begin{figure*}[!ht]
\centering
\includegraphics[clip=true,trim= 0 0 0 0, width=1\textwidth]{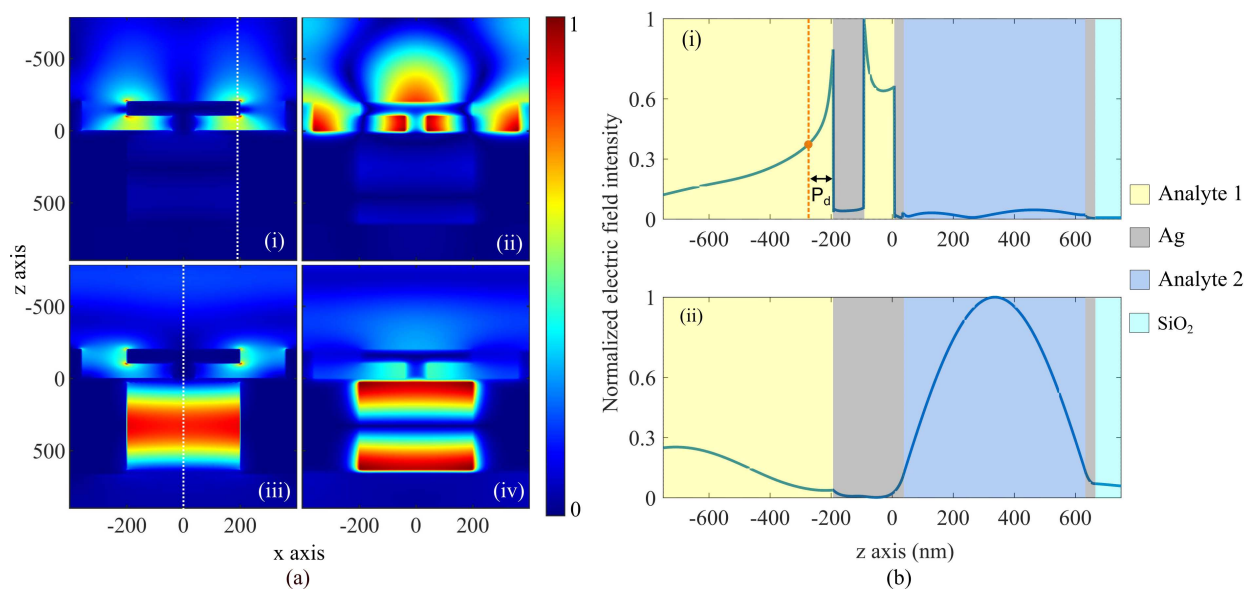}
\caption{\label{fig:fields} \textcolor{black}{Normalized (a,i) electric and (a,ii) magnetic field profile at resonant wavelength of 1119 nm (mode 1). Similarly, (a,iii) electric and (a,iv) magnetic field profile at resonant wavelength of 1847 nm (mode 2). (b,i) The normalized line plot of electric field intensity along x = 195 nm (depicted in white dotted line) of (a,i). Here, the penetration depth (P\textsubscript{d}) is found to be 81.63 nm outside the grating surface depicted by an orange dotted line. Likewise, (b,ii) the normalized line plot of electric field intensity along x = 0 nm (depicted in white dotted line) of (a,iii). For all the cases, the structure is illuminated by a TM polarized EM wave along the negative z axis.}}
\end{figure*}

\begin{figure*}[!h]
\centering\includegraphics[clip=true,trim= 0 0 0 0,width=0.85\linewidth]{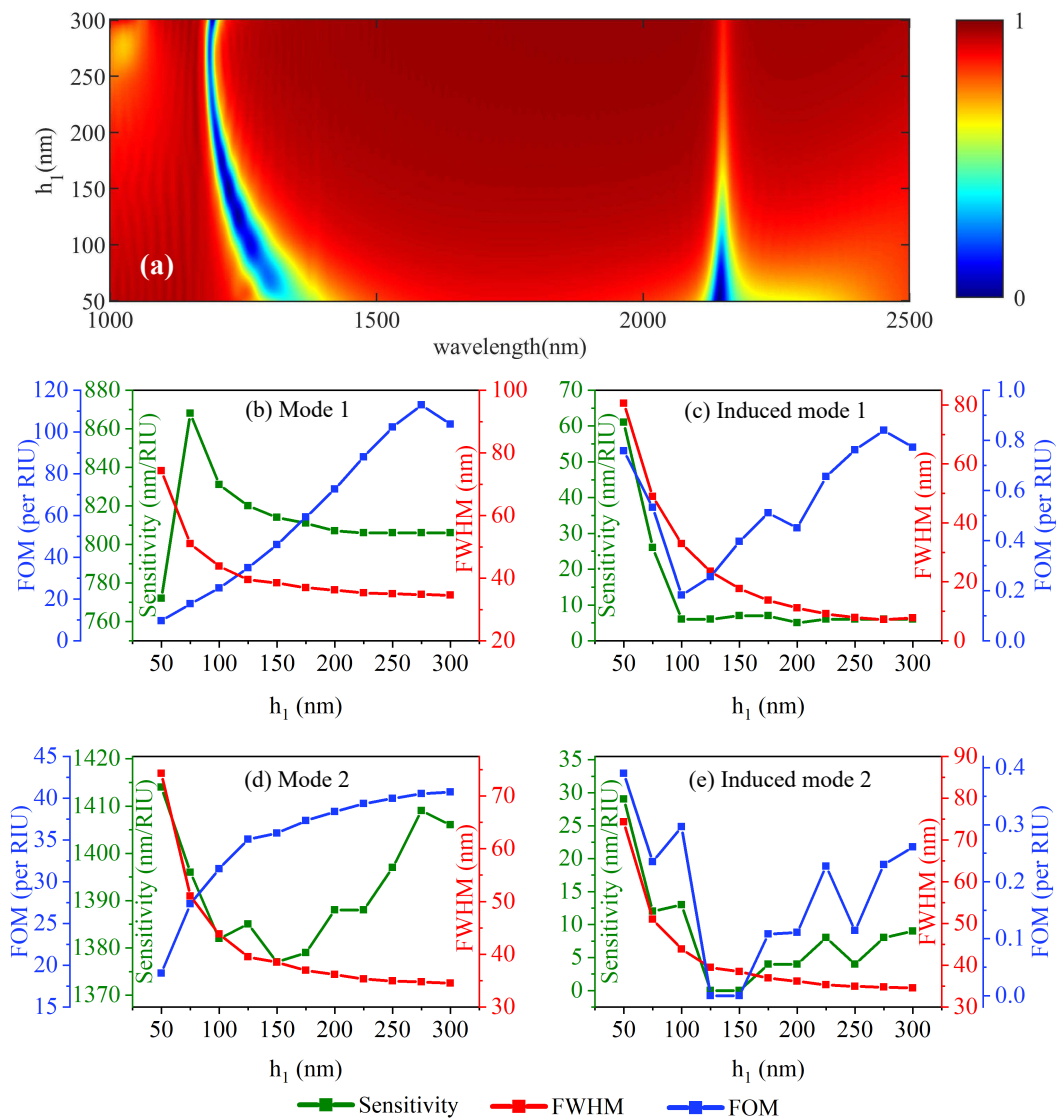}
\caption{\label{fig:h1} (a) Reflection spectrum for different h\textsubscript{1} values from 50 nm to 300 nm while mode 1 manifests a spectral blueshift and mode 2 resonance is almost fixed. (b) Individual and (c) induced sensing parameters of mode 1 where individual performance improves with h\textsubscript{1}.  (d) Individual and (e) induced sensing parameters of mode 2. Similar to mode 1, the individual performance of mode 2 improves with h\textsubscript{1}.}
\end{figure*}

\textcolor{black}{Field distributions from fig. \ref{fig:fields} give us an analogy behind the individuality of the two resonance modes. The electric field and magnetic field profile of mode 1 in figs. \ref{fig:fields}(a,i) and \ref{fig:fields}(a,ii) show high confinement of electric and magnetic fields around the L-shaped cavity (region 1) due to trapped photon-electron interaction while almost zero electromagnetic field perturbation in region 2, enabling a low induced spectral response of mode 2 for a change in RI of region 1. On the contrary, the field distributions of mode 2 in figs. \ref{fig:fields}(a,iii) and \ref{fig:fields}(a,iv) depict confined electromagnetic field in region 2 but due to the off-resonant electromagnetic field distribution around the L-shaped cavity, a relatively higher induced spectral response of mode 1 (6.5 nm RIU\textsuperscript{-1}) is observed than that of mode 2 (2 nm RIU\textsuperscript{-1}).}

\textcolor{black}{To have a deeper understanding of the distinctive resonant characteristics of the sensor, a 1D line plot of corresponding normalized electric field distribution from fig. \ref{fig:fields}(a,i) at x = 195 nm is illustrated in fig. \ref{fig:fields}(b,i). Interestingly, for mode 1 resonance, the electric field is highly confined within the L-shaped cavity and outside the grating region, the evanescent field decays to its $\sim$37\% of the maximum value at the interface within 81.63 nm (penetration depth, P\textsubscript{d}) of the analyte 1 region outside the grating while only 4.6\% of the maximum electric field is confined in region 2. Similarly, for mode 2 resonance, the 1D line plot corresponding to the electric field distribution of fig. \ref{fig:fields}(a,iii) at x = 0 nm is depicted in fig. \ref{fig:fields}(b,ii). Notably, the electric field is confined to region 2 and the field intensity is in good agreement with previously reported intensity profiles within FP cavities, which further confirms mode 2 as an FP-like resonance \cite{neumeier2015self,kang2011enhanced}. Here, for both resonant modes, the field intensity distribution is highly confined to the corresponding region of analytes conveying the potential inter-region interference suppressing capability of the proposed sensor structure.}

It is to be noted that the induced sensitivity for a broad spectral region of interest introduces some error in the individual operation of the resonance modes but for our proposed sensor this induced error is significantly lower than previously reported simultaneous multi-analyte and self-referenced sensors \cite{Sharma2019,Tabassum2020,Abutoama2015}. It is possible to reduce the induced error even further by optimizing the structural parameters of the T-shaped grating.

\begin{figure*}[!ht]
\centering
\includegraphics[clip=true,trim= 0 0 0 0,width=0.8\linewidth]{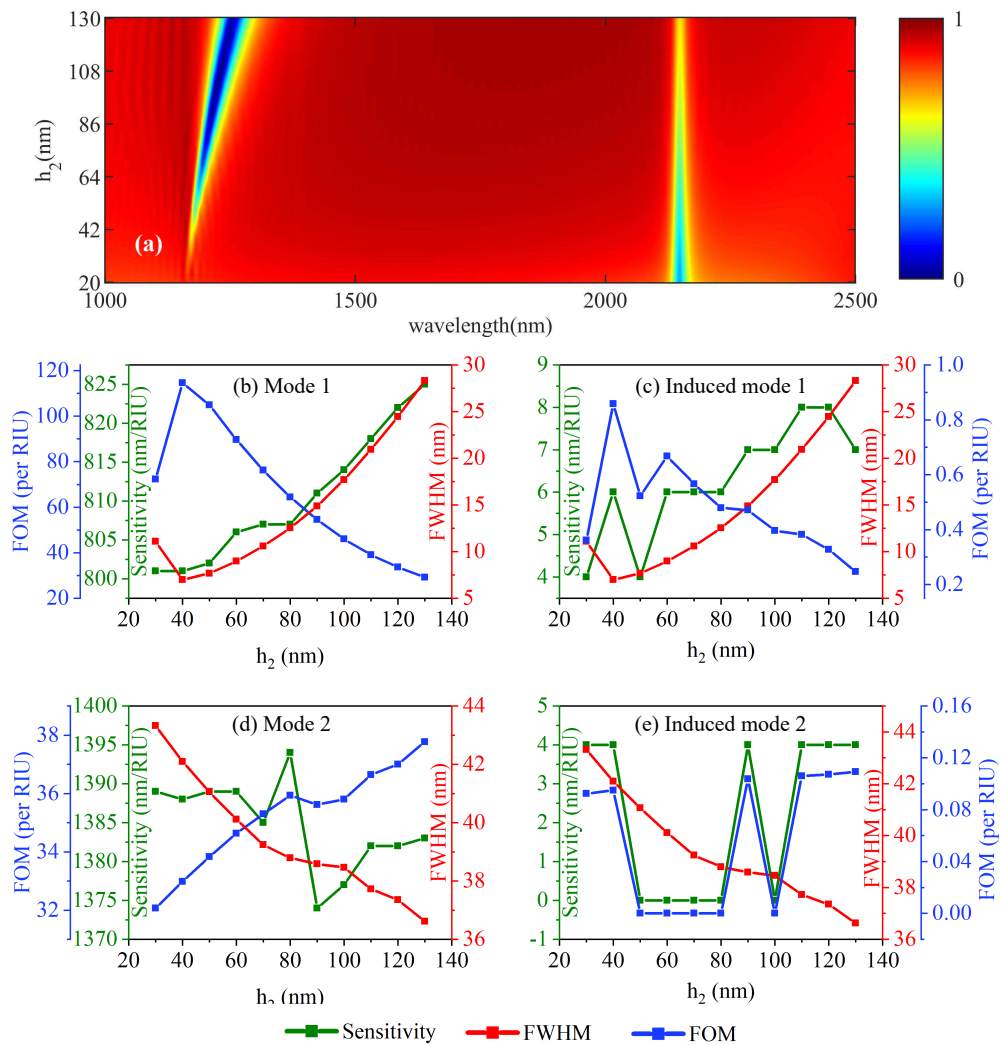}
\caption{\label{fig:h2} (a) Reflection spectrum for different h\textsubscript{2} ranging from 30 nm to 150 nm. Mode 1 occurs when h\textsubscript{2} is greater than 30 nm and undergoes a small redshift with the increase of h\textsubscript{2} while mode 2 does not undergo any significant spectral shift. (b) Individual and (c) induced sensing parameters of mode 1, (d) individual and (e) induced sensing parameters of mode 2 for various values of h\textsubscript{2}.}
\end{figure*}

\subsection{Characterization of T-shaped grating parameters:}

Optimizing the T-shaped grating parameters can provide a better spectral response while providing lower induced sensitivities for the two separate modes. To study the influence of grating parameters on overall sensing performance, the grating parameters have been swept each at a time while keeping others fixed. During optimization, the RI of each region is considered 1.33. At first, the effect of h\textsubscript{1} on the reflection spectra is observed. In fig.  \ref{fig:h1}(a), the resonant wavelength of mode 1 undergoes blueshift as h\textsubscript{1} increases and the strong field confinement occurs at h\textsubscript{1} = 150 nm. As increasing cavity height results in increasing the cavity volume, higher photon energy is required for the resonance, so the blueshift occurs. On the contrary, mode 2 does not experience any significant spectral shift but the resonance weakens as h\textsubscript{1} increases. \textcolor{black}{From fig. \ref{fig:h1}(b)-(c) it can be seen that with the increase of h\textsubscript{1} the FWHM decreases which means that with the increase of the cavity height, narrow-band coupling occurs and a less electric field is confined. However, for mode 2, the decrease of the FWHM shown in fig. \ref{fig:h1}(d)-(e) and can be explained by the reflectance dependence of the FP cavity modes\cite{born2013principles,kitchin2003astrophysical}:
\begin{equation}
      \lambda_{FWHM} \propto \frac{1-r}{\sqrt{r}}
\end{equation}
Where $\lambda$\textsubscript{FWHM} is the FWHM of the FP-like cavity mode and r is the reflectance of that mode.} The induced sensitivity for mode 1 in fig. \ref{fig:h1}(c) drops from 60 nm RIU\textsuperscript{-1} to 6 nm RIU\textsuperscript{-1} when h\textsubscript{1} increases from 50 nm to 100 nm and remains fixed beyond h\textsubscript{1} = 100 nm. On the contrary, the induced sensitivities of mode 2 in fig. \ref{fig:h1}(e)  does not follow any trend with the increase of h\textsubscript{1} but at h\textsubscript{1} = 125 nm and 150 nm the mode 2 induced sensitivity drops to 0 nm RIU\textsuperscript{-1}. At h\textsubscript{1} = 150 nm, the lowest induced mode 2 sensitivity can be achieved while other individual sensing parameters are high enough to provide a better overall sensing performance. 

\begin{figure*}[!h]
\centering
\includegraphics[clip=true,trim= 0 0 0 0,width=0.85\linewidth]{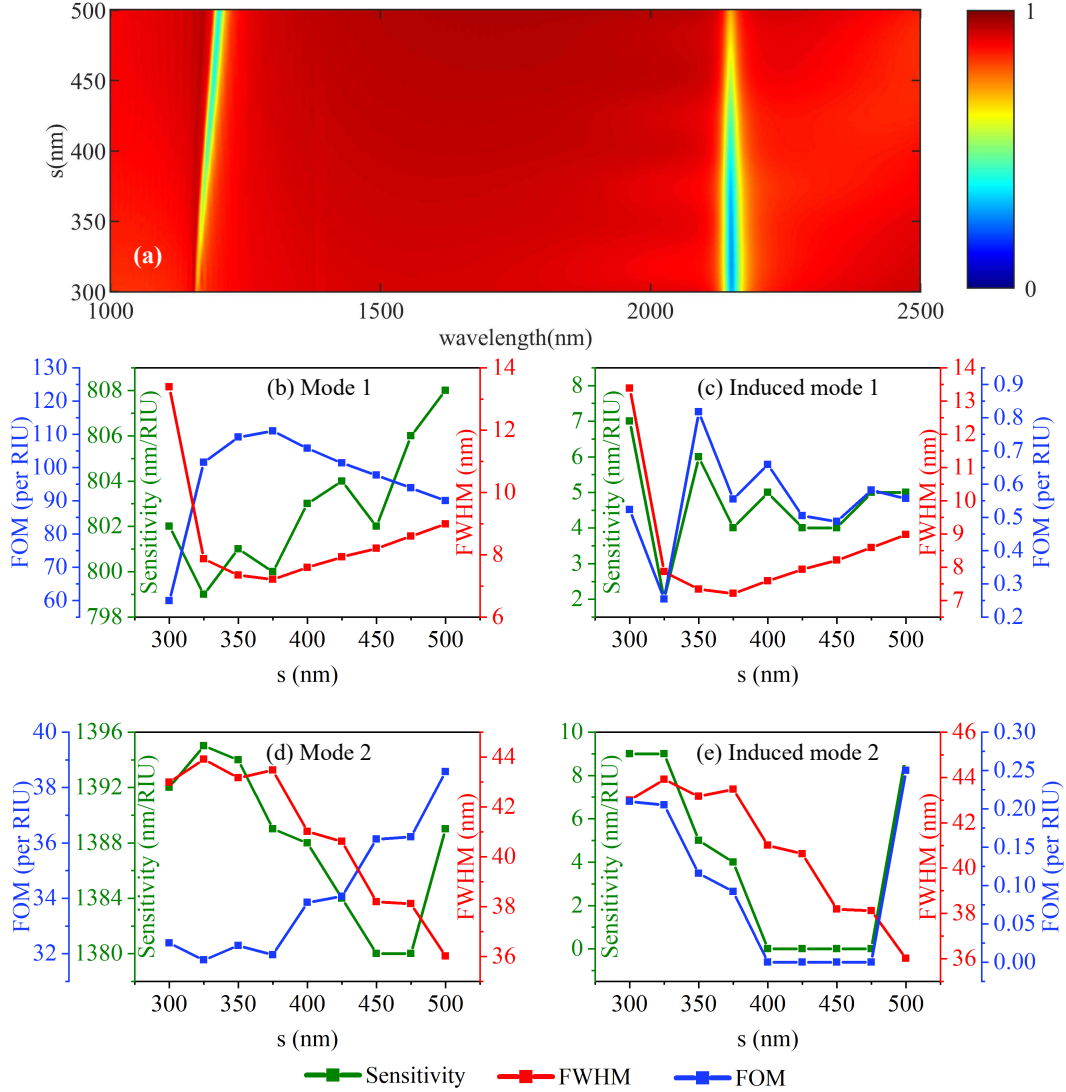}
\caption{\label{fig:s}(a) Reflection spectrum for different values of s ranging from 300 nm to 500 nm. With the increase of s, mode 1 resonance becomes more pronounced while mode 2 resonance gradually weakens. (b) Individual and (c) induced sensing parameters of mode 1 for various values of s. The individual FOM of mode 1 increases with s and beyond s = 375 nm, the FOM decreases gradually. (d) Individual and (e) induced sensing parameters of mode 2 where individual FOM gradually improves with s.}
\end{figure*} 

The reflection spectrum for the various h\textsubscript{2} is plotted in fig. \ref{fig:h2}(a). For h\textsubscript{2} < 30 nm mode 1 resonance does not occur as the horizontal slab thickness becomes smaller than the skin depth of Ag and most of the incident electric field penetrates through the slab and strongly couples to region 2. At h\textsubscript{2} > 30 nm, a strong coupling has been observed with the increase of h\textsubscript{2} while mode 2 resonance weakens gradually as most of the electric field confines at region 1. Comparing with h\textsubscript{1}, horizontal axis thickness (h\textsubscript{2}) has considerably less significant impact on shifting the mode 1 resonant wavelength.\textcolor{black}{While h\textsubscript{2} increases as depicted in fig. \ref{fig:h2}(b)-(c), the electric field of mode 1 couples more strongly in the horizontal slab, and thus energy dissipation increases causing a wider FWHM profile for mode 1. As mode 1 exhibits strong coupling with the increasing h\textsubscript{2}, mode 2 weakens rendering low energy dissipation and narrower FWHM profile as seen in fig. \ref{fig:h2}(d)-(e).} 

From fig. \ref{fig:h2}(b), with the increase of h\textsubscript{2}, individual FOM of mode 1 increases up to h\textsubscript{2} = 40 nm then gradually decreases while individual FOM of mode 2 in fig. \ref{fig:h2}(d) gradually improves. The induced sensitivity for both mode 1 (fig. \ref{fig:h2}(c)) and mode 2 (fig. \ref{fig:h2}(e)) does not follow any significant trend with the increase of h\textsubscript{2} but at h\textsubscript{2} = 50 nm induced sensitivity of mode 2 is 0 nm RIU\textsuperscript{-1} while induced mode 1 sensitivity is 4 nm RIU\textsuperscript{-1} with high individual sensing parameters of each mode.

The effect of various horizontal slab span (s) on \textcolor{black}{the reflection spectrum} in fig. \ref{fig:s}(a) shows that with the increase of s the surface area for SPR coupling increases which results in better confinement of electric and magnetic field, so the resonance for mode 1 strengthens. As the total energy is constant, strong coupling at mode 1 results in weak coupling at mode 2.  The effect of s is less significant on the resonant wavelength compared to h\textsubscript{1} and even h\textsubscript{2}. \textcolor{black}{As shown in fig. \ref{fig:s}(b)-(c), for s>370 nm the FWHM increases for mode 1 which is because increased value of s results in narrowing the area between the ridge of the horizontal slab and neighbouring wall. So, the electric field confinement and energy dissipation increases and thus, the FWHM broadens. On the contrary, from fig. \ref{fig:s}(d)-(e), the FWHM of mode 2 decreases eventually with the increase of s which infers that electric field confinement reduces with the increasing s.} 

In fig. \ref{fig:s}(b), FOM for individual operation of mode 1 increases with s up to s = 375 nm and then decreases while mode 2 FOM in fig. \ref{fig:s}(d),  gradually increases. The induced sensitivity of mode 2 is 0 nm RIU\textsuperscript{-1} from s = 400 nm to s = 475 nm as depicted in fig. \ref{fig:s}(e). In terms of both induced and individual sensing parameters, overall optimum performance can be achieved when s = 400 nm.

\begin{figure}
\centering
\includegraphics[clip=true,trim=0 0 0 0, width=1\textwidth]{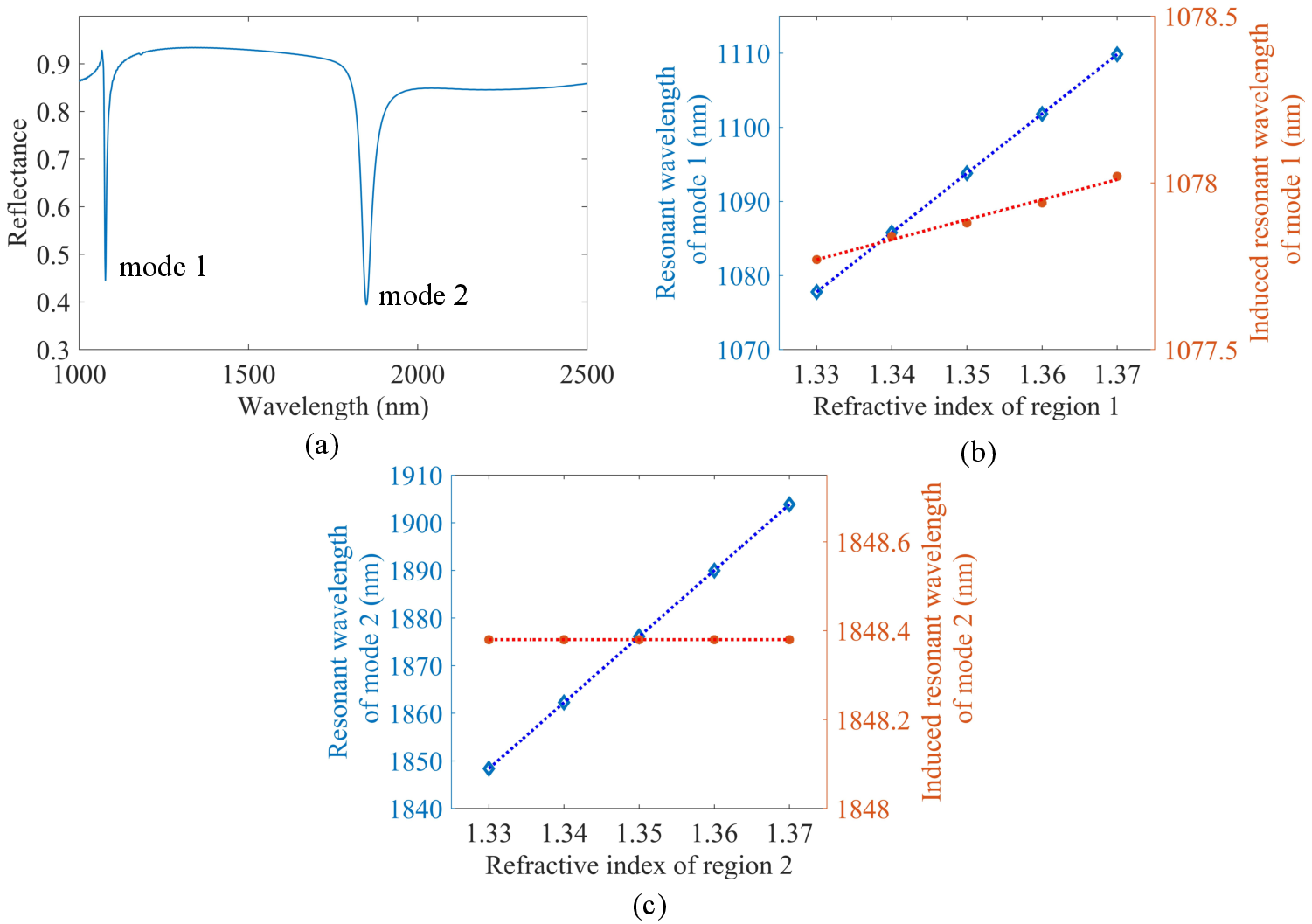}
\caption{\label{fig:reso} (a) Reflection spectrum of the sensor with h\textsubscript{1} = 150 nm, h\textsubscript{2} = 50 nm, s = 400 nm, t\textsubscript{1} = 80 nm and t\textsubscript{2} = 80 nm. The RI of both regions was set to 1.33. FWHMs of mode 1 and mode 2 are 7.78 nm and 41.29 nm respectively. Resonant wavelength for both individual and induced performance of (b) mode 1 and (c) mode 2.}
\end{figure}

\subsection{Optical response with optimum grating parameters:}

According to the previous discussion, the T-shaped grating parameters have been studied to find the optimum dimensions of the proposed sensor for further reduction of induced sensitivities with better individual sensing performance and reasonable electric field confinement in both regions. Fig. \ref{fig:reso}(a) is showing the reflection spectrum of the optimized sensor with h\textsubscript{1} = 150 nm, h\textsubscript{2} = 50 nm, s = 400 nm, t\textsubscript{1} = 80 nm and t\textsubscript{2} = 80 nm. The reflection spectrum has been plotted considering 1.33 as the RI of both analytes where the FWHMs 7.78 nm and 41.29 nm have been recorded for mode 1 and mode 2 respectively which also improve overall FOM of the proposed sensor.

Furthermore, the resonant modes hold following linear relations with the RI of both regions as plotted in fig. \ref{fig:reso}(b) and \ref{fig:reso}(c).
\begin{equation}
\begin{aligned}\label{eq:simul}
\lambda_{11} &= 801.7 \times n_1 + 11.515  \\
\lambda_{12} &= 5.80 \times n_2 + 1070.062   \\
\lambda_{22} &= 1386.8 \times n_2 + 3.942  \\
\lambda_{21} &= 7.4002 \times 10^{-12}\times n_1+ 1848.38 
\end{aligned}
\end{equation}

\begin{figure}[h]
\centering\includegraphics[clip=true,trim=0 0 0 0, width=1\textwidth]{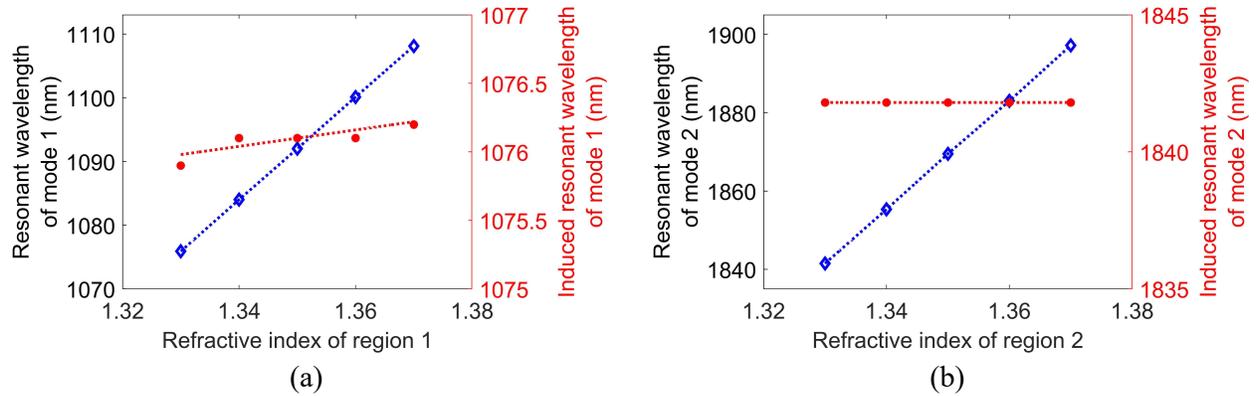}
\caption{\label{fig:for_au} Individual and induced performance of (a) mode 1 and (b) mode 2 resonance for different refractive index ranging from 1.33 to 1.37 when the silver is replaced with gold in the structure.}
\end{figure}

Here, $\lambda_{11}$ ($\lambda_{22}$) denotes the resonant wavelength for individual operation of mode 1 (mode 2) and $\lambda_{12}$ ($\lambda_{21}$) is the induced resonant wavelength of mode 1 (mode 2) when RI of region 1 is fixed at 1.33 and n{\textsubscript{1}} (n{\textsubscript{2}}) is the RI of region 1 (region 2). When the refractive indices of the regions change simultaneously, the resonant wavelength of each mode includes both of its individual and induced performance. Hence,
\begin{equation}
\begin{aligned}\label{eq:res}
\lambda_{1} &= \lambda_{11} + \lambda_{12}  \\
\lambda_{2} &= \lambda_{22} + \lambda_{21} 
\end{aligned}
\end{equation}
From Eqs. (\ref{eq:simul}) and (\ref{eq:res}), following expression can be derived:
\begin{equation}\label{eq:error}
\begin{aligned}
     \lambda_1 = 801.70&\times n_1 + 4.18\times 10^{-3}\times \lambda_{2} +1080.234 \\
     \lambda_2 &= 1386.80 \times n_2 + 1852.32
\end{aligned}
\end{equation}
Here, Eq. (\ref{eq:error}) can be used to detect the simultaneous change in RI of both regions mitigating the error introduced by induced sensitivity. From Eq. (\ref{eq:fom}), the calculated FOMs of mode 1 and mode 2 are 103.05 RIU\textsuperscript{-1} and 33.58 RIU\textsuperscript{-1} respectively. Moreover, the refractive index sensing performance of the sensor incorporating gold (palik)\cite{palik1998handbook}, instead of silver was investigated. The variation of resonant wavelength position with refractive of the corresponding regions are depicted in fig. \ref{fig:for_au}(a) and \ref{fig:for_au}(b) for mode 1 and mode 2 respectively. Interestingly, we find analogous results for gold where mode 2 exhibits no induced resonant wavelength shift due to the corresponding refractive index variation (fig. \ref{fig:for_au}(b)). Thus, we can justify the proposition of our structure to minimize the induced sensitivity and accurately detect the refractive index of both analytes simultaneously.     
Furthermore, both the individual and induced resonant modes hold the following linear relations with the RI of the corresponding regions.
\begin{equation}
\begin{aligned}\label{eq:simul_Au}
\lambda_{11} &= 805 \times n_1 + 5.27  \\
\lambda_{12} &= 6.00 \times n_2 + 1068   \\
\lambda_{22} &= 1391 \times n_2 - 8.55  \\
\lambda_{21} &= 9 \times 10^{-12}\times n_1+ 1841.8 
\end{aligned}
\end{equation}
Employing the similar calculations, the error in refractive index calculation can also be minimized by using the following relations.
\begin{equation}
    \label{eq:error_Au}
\begin{aligned}
    &\lambda_1 = 805\times n_1 + 4.313\times 10^{-3}\times \lambda_{2} + 1071.95 \\
    &\lambda_2 = 1391\times n_2 + 1833.25
\end{aligned}
\end{equation}
Here the variables in  denotes similar parameters described earlier. As the two sets of results are in well agreement with each other, the accuracy of the resonant modes for refractive index ranging from 1.33 to 1.37 can also be achieved for gold as a structural material of the proposed sensor. The simultaneous sensing performance comparison of the proposed sensor where Ag is the structural material with other previously reported sensors is included in table \ref{tab:tab1}. 


\begin{table*} [b]
    \centering
     \caption{Performance comparison of the proposed sensor with reported double resonance sensors}
   \newcolumntype{x}[1]{>{\centering\arraybackslash\hspace{0pt}}p{#1}}
   \begin{tabular}{m{4cm} x{2cm} x{2cm} x{2cm} x{2cm} x{2cm}}
    \toprule
    
       Structure & \parbox[b]{\hsize}{\centering Sensitivity (nm/RIU)} & FWHM (nm) & FOM (RIU\textsuperscript{-1}) & RI range & Reference  \\ 
        \midrule
         \makecell[l]{Bragg grating based gold \\  coated multi-channel fiber}  & 200 & 0.285 & 700 & -- & \cite{Spackova2009}\\ 
         \hline
         \makecell[l]{Cascaded single-mode\\ fiber (SMF)  with \\ thin-core fiber (TCF) } & 30.29 and 79.335 & -- & -- & 1.333$\sim$1.3786 & \cite{Sun2018}\\ 
         \hline
         \makecell[l]{Side coupled metal-\\dielectric-metal slot cavity} & 1131 & $7.06\times10^{-5}$ & $1.6\times10^{7}$ & 1.00$\sim$1.10 & \cite{Wen2016}\\ 
         \hline
         \makecell[l]{Cross rectangular cavity with \\ metal-insulator-metal (MIM)}
 & 980 and 1040 & 4.95 and 5.22 & 197.6 and 198.9 & -- & \cite{Fu2018}\\ 
 \hline
 \makecell[l]{Polymeric micro-ring resonator}
& 200 & 0.15 & 1333.33 & -- & \cite{Kim2008} \\ 
\hline
\makecell[l]{Whispering gallery resonator
}& 240 & 1 & 240 & -- & \cite{Chen2019} \\ 
\hline
\makecell[l]{T-shaped grating on top of \\ periodic nano-cavities
}& 801.7 and 1386.8 & 7.78 and 41.29  & 103.05 and 33.58 & 1.33$\sim$1.37 & this work \\
      \bottomrule
    \end{tabular}
    \label{tab:tab1}
\end{table*}

\section{Performance evaluation of the sensor}
\begin{figure}
\centering\includegraphics[clip=true,trim=0 0 0 0, width=1\textwidth]{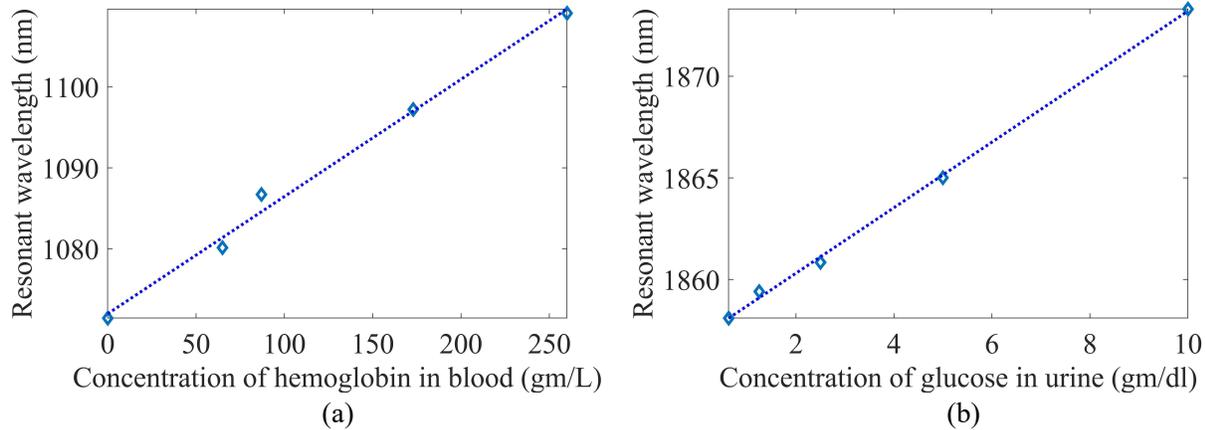}
\caption{\label{fig:multi} Resonant wavelengths for the concentration of (a) hemoglobin in blood (region 1) (b) glucose in urine (region 2)}
\end{figure}

The capability of the structure to perform as a biosensor has been studied under three different conditions: (i) simultaneous sensing of two analytes at two different regions, (ii) more reliable sensing of one analyte and (iii) self-referencing detection to eliminate environmental noises. Following performance verification was done with the optimum grating structure with h\textsubscript{1} = 150 nm, h\textsubscript{2} = 50 nm and s = 400 nm. 
\subsection{Simultaneous multi-analyte detection}
Hemoglobin is an iron-containing protein that resides in red blood cells. It binds oxygen from the lungs and delivers it throughout the body by red blood cells. Insufficient hemoglobin causes deficiency of oxygen inside the body and the condition is known as anemia. A low level of hemoglobin is also an indicator of other complications: iron-deficiency, unusual hemolysis (breaking down of red blood cells) or vitamin deficiency. Pregnant women and newborn infants may suffer from anemia. Low blood oxygen level may have the symptoms of weakness, headache, dizziness, chest pain, pale skin and shortness of breath. Again, a high level of hemoglobin indicates polycythemia i.e. denser blood than usual which can cause blood clots, strokes and heart attacks. So, the concentration of hemoglobin should be monitored reliably. The RI of hemoglobin in an aqueous solution changes linearly with the concentration at room temperature and can be fitted with the following equation where $n_h$ and $c_h$ denote RI of the solution and concentration of hemoglobin in it respectively \cite{Lazareva2018}. 
\begin{equation}
    n_h = 0.00018064 \times c_h + 1.3227
\end{equation}

Diabetes is a chronic disease that affects the metabolic system of the body. Diabetic patients suffer from hyperglycemia (high blood sugar) as the body cannot produce the required insulin. A long-lasting excess sugar concentration in blood may cause heart diseases, kidney complications and vision loss. Diabetic patients are needed to be under regular supervision for blood sugar level. Excess glucose is spilled into the urine causing frequent urination. So, blood glucose can be monitored by diagnosing urine sugar level. RI of urine ($n_g$) changes almost linearly with the concentration ($c_g$) of dissolved glucose\cite{Ahmad2010} and can be expressed by the following equation:
\begin{equation}
    n_g = 0.0012 \times c_g + 1.3363 \\
\end{equation}

For the simultaneous detection of the concentrations of hemoglobin and glucose, the RI of region 1 was altered according to the different concentrations of hemoglobin while the RI of region 2 was varied for corresponding concentrations of glucose in urine. Here is to be noted that, the diameter of a single molecule of hemoglobin is 5 nm \cite{Erickson2009} which is 30 times smaller than the L-shaped cavity height under T-shaped grating, so the assumption of homogeneous concentration for bulk RI sensing is valid. The simulated results for hemoglobin and glucose are depicted in Fig. \ref{fig:multi}(a) and \ref{fig:multi}(b) respectively where resonant wavelengths hold the following linear relations:
\begin{equation}
\begin{aligned}
   \lambda_1 &= 0.14491\times c_h + 1071.9639 \\
  \lambda_2 &= 1.6125\times c_g + 1857.0854
\end{aligned}
\end{equation}

With a typical optical spectrum analyzer (OSA) having a spectral resolution of 0.05 nm \cite{hui2009fiber}, the proposed sensor can detect a hemoglobin concentration as low as 0.03 gm per deciliter using it as region 1 analyte while the average concentration of hemoglobin for a middle-aged human is 13 to 15 gm/deciliter \cite{HAWKINS1954}. Using as region 2 analyte, minimum 0.03 gm/dl or 0.17 mmol/dl urine glucose can be detected while the accuracy range for commercial glucometers is ± 83 mmol/dl. 
\begin{figure}[t]
\centering\includegraphics[clip=true,trim=0 0 0 0, width=0.5\textwidth]{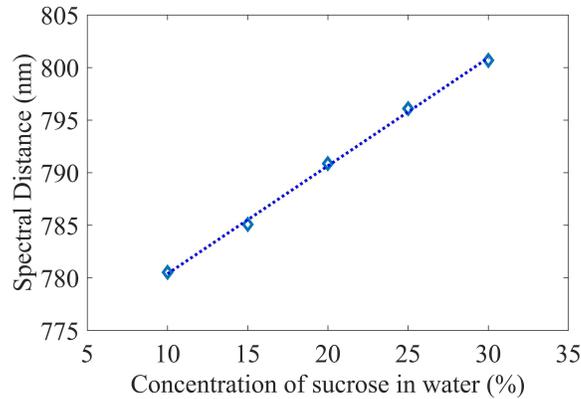}
\caption{\label{fig:sucrose} Spectral distance between two resonance modes for different concentrations of sucrose in water. }
\end{figure}

\subsection{Improving accuracy}

Sensing the RI of the same analyte simultaneously with two resonant modes provides improved accuracy. With the increase in RI of the same surrounding material, the resonant modes undergo a redshift, but for having different sensitivity, each mode shifts at a different but specific rate resulting in broadening the spectral distance between them. Considering this spectral distance shift for different RI of the surrounding medium as a sensing parameter, two-stage verification, hence more reliable performance can be harnessed. To investigate the reliable sensing performance, the proposed structure has been studied to detect sucrose concentration in an aqueous medium. The refractive index data for five different concentrations of sucrose has been derived from a reported literature \cite{hwang1994experiments}. As seen in fig. \ref{fig:sucrose}, the relation between spectral distance and sucrose concentration holds a linear relation depicted by:
\begin{equation}
    p = 1.0262 \times c_s+ 770.126
\end{equation}
Here, $p$ is the spectral distance between the resonant modes in nm and $c_s$ is the percentage of sucrose in the solution. This sensor can reliably detect $6.07 \times 10^{-6}$ \% change in the concentration of sucrose in an aqueous medium.

\subsection{Self-reference mode}
\begin{figure}[b]
\centering\includegraphics[clip=true,trim=0 0 0 0, width=0.5\textwidth]{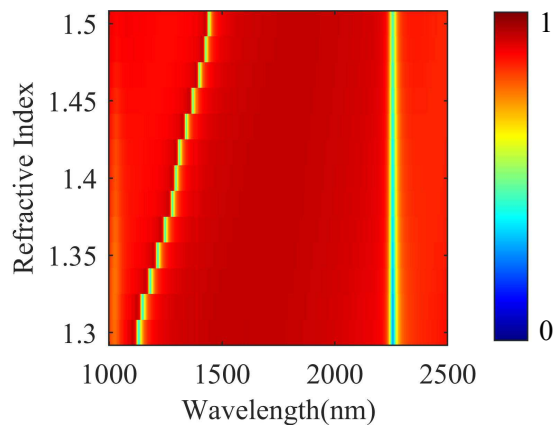}
\caption{\label{fig:self} Reflection spectra of the sensor for different refractive indices of region 1 while region 2 was filled with SiO\textsubscript{2}}
\end{figure}
While optical refractive index sensors are powerful and efficient tools for bio-chemical sensing and binding event monitoring \cite{Augel2018}, being highly sensitive to the local index of the surrounding medium of the metal surface introduces heavy environmental noise in their optical response due to light intensity fluctuation \cite{Wang2017self}, temperature variation \cite{Homola1999}, the humidity of the unstable environment \cite{Montero2009}. However, these evaluation difficulties can be avoided by coupling electromagnetic fields in different regions of a sensor structure at different phase-matching conditions where one region is physically separated from the environment. Thus a self-referenced refractive index sensor can be designed and fabricated which is immune to the unsteady environmental parameters. As discussed earlier, in our proposed sensor the mode 2 couples to region 2 which is physically separated from region 1 by a thin silver film, hence utilizing this unique property, the sensor has enormous potential as a self-referenced RI sensor.

To explore the self-referencing capability of the sensor, a set of simulations has been executed under different refractive indices ranging from 1.3 to 1.5 for region 1 while region 2 was filled with SiO\textsubscript{2}. As seen in fig. \ref{fig:self}, under this broad range of refractive index, the resonant wavelength of mode 2 remains almost fixed with some insignificant fluctuation while mode 1 resonant wavelength undergoes a redshift which is expected because of the dielectric loading \cite{shahbazyan2013plasmonics}. The self-referencing performance comparison of the proposed sensor with other previously reported self-referencing sensor is included in table \ref{tab:tab2}.

\begin{table*} [!h]
    \centering
     \caption{Performance comparison of the proposed self-referenced sensors}
   \newcolumntype{x}[1]{>{\centering\arraybackslash\hspace{0pt}}p{#1}}
   \begin{tabular}{m{3.8cm} x{1.8cm} x{1.8cm} x{1.8cm} x{1.5cm} x{1.3cm} x{1.2cm}}
    
    \toprule
Structure                                                   
&
\parbox[b]{\hsize}{\centering Active Mode Sensitivity (nm/RIU)} & 
\parbox[b]{\hsize}{\centering Ref. Mode Sensitivity (nm/RIU)} 
& \parbox[b]{\hsize}{\centering Active Mode FWHM (nm)}
& \parbox[b]{\hsize}{\centering Active Mode FOM (RIU\textsuperscript{-1})}
& RI Range 
& Reference \\
\midrule 

 \makecell[l]{Titanium   oxide (TiO2) \\ grating on thin gold (Au) \\ film} & 693.88                                              &  23.33 & 27.5 & 25.23                         & 1.32$\sim$1.35         & \cite{Sharma2019}  \\ 
 \hline
\makecell[l]{Ag   grating on top of \\ SiO2 substrate}                                         & 445                                                 & -- & -- & -- & 1.329$\sim$1.359       & \cite{Karabchevsky2009} \\ 
\hline
\makecell[l]{Si\textsubscript{3}N\textsubscript{4} on   top of Ag thin film \\ on SiO2 substrate}                              & 580                                                 & 40  & -- & -- & 1.33$\sim$1.34         & \cite{Abutoama2015}   \\
\hline
\makecell[l]{U-shaped   cavities side-\\ coupled to a metal–\\dielectric–metal (MDM) \\ waveguide}  & 917                                                 & --   &  5.09 & 180  & 1.000$\sim$1.015       & \cite{Ren2018}      \\
\hline
\\Metallic   grating structure                                                  & 470                                                 & --                     & 15.16 & 31 & 1.33$\sim$1.35         & \cite{wang2017}        \\ 
\hline
\\T-shaped grating on nano-cavity                                               & 800                                                 & 2.82                                                   & 7.78 & 103.05 & 1.3$\sim$1.5          & this work \\
\bottomrule                             
\end{tabular}
\label{tab:tab2}
\end{table*}


\section{Conclusion}
Refractive index sensor with multi-analyte sensing feature has always been a center of research interest. In this paper, a refractive index sensor based on a novel grating structure on top of periodic nano-cavities has been proposed and numerically analyzed. Due to the generation of two individual resonance modes originating from SPR and \textcolor{black}{FP-like resonant mode} respectively, a phenomenal sensing performance with insignificant inter-region interference has been found for a broad refractive index window ranging from 1.3 to 1.5 which is larger than most of the similar reported sensors. Average sensitivities for individual performance were found as high as 801.7 nm RIU\textsuperscript{-1} and 1386.8 nm RIU\textsuperscript{-1} respectively. The potential application of the sensor for multi-analyte and self-referenced sensing have been numerically verified. Due to its versatility, we believe, the proposed sensor is a prominent candidate for biochemical sensing application. 

\section*{Funding}
BUET CASR Research Projects Under Higher Training \& Research Programme (CASR Meeting No.: 339, Resolution No.: 62, Meeting Date: 07/04/2021)

\section*{Disclosure}
The authors declare no conflicts of interest.

\section*{Data Availability Statement}
Data underlying the results presented in this paper are not publicly available at this time but may be obtained from the authors upon reasonable request.

\typeout{}
\bibliographystyle{unsrt}  
\bibliography{references}  


\end{document}